\documentstyle[12pt,epsf]{article}

\newlength{\pubnumber} 

\catcode`\@=11
\@addtoreset{equation}{section}

\def\section{\@startsection{section}{1}{\z@}{3.5ex plus 1ex minus .2ex}
 {2.3ex plus .2ex}{\large\bf}}
\def\subsection{\@startsection{subsection}{2}{\z@}{2.3ex plus .2ex}
 {2.3ex plus .2ex}{\bf}}

\begin{document}

\begin{titlepage}
\samepage{
\setcounter{page}{1}
\rightline{ACT-04/00}
\rightline{CTP-TAMU-07/00}
\rightline{OUTP--00--04P}
\rightline{TPI-MINN-00/8}
\rightline{UMN--TH--1843-00}
\rightline{\tt hep-ph/0002292}
\rightline{February 2000}
\vfill
\begin{center}
 {\Large \bf Towards String Predictions}
\vfill
\vskip .4truecm
\vfill {\large
        G.B.~Cleaver,$^{1,2}$\footnote{gcleaver@rainbow.physics.tamu.edu}
        A.E.~Faraggi,$^{3,4}$\footnote{faraggi@mnhepw.hep.umn.edu}
        D.V.~Nanopoulos$^{1,2,5}$\footnote{dimitri@soda.physics.tamu.edu},}

       {\large and
        T.~ter Veldhuis$^{3}$\footnote{veldhuis@hep.umn.edu}}
\\
\vspace{.12in}
{\it $^{1}$ Center for Theoretical Physics,
            Dept.\  of Physics, Texas A\&M University,\\
            College Station, TX 77843, USA\\}
\vspace{.025in}
{\it $^{2}$ Astro Particle Physics Group,
            Houston Advanced Research Center (HARC),\\
            The Mitchell Campus,
            Woodlands, TX 77381, USA\\}
\vspace{.025in}
{\it$^{3}$  Department of Physics, University of Minnesota, 
            Minneapolis, MN 55455, USA,\\}
\vspace{.025in}
{\it$^{4}$ 	 Theoretical Physics, University of Oxford, 1 Keble Road,
Oxford OX1 3NP, UK\\}
\vspace{.025in}
{\it$^{5}$  Academy of Athens, Chair of Theoretical Physics, 
            Division of Natural Sciences,\\
            28 Panepistimiou Avenue, Athens 10679, Greece\\}
\vspace{.025in}
\end{center}
\vfill
\begin{abstract}
The aim of superstring phenomenology is to develop the tools and
methodology needed to confront string theory with experimental data.
The first mandatory task is to find string solutions which
reproduce the observable data. The subsequent goal
is to extract potential signatures beyond the observable data. 
Recently, by studying exact flat directions of non--Abelian singlet fields,
we demonstrated the existence of free fermionic heterotic--string models
in which the $SU(3)\times SU(2)\times U(1)_Y$--charged matter spectrum,
just below the string scale, consists solely of the MSSM spectrum. 
In this paper we study the possibility that the exact flat directions
leave a $U(1)_{Z^\prime}$ symmetry unbroken at the Planck scale.
We demonstrate in a specific example that such unbroken $U(1)_{Z^\prime}$
is in general expected to be not of the GUT type but
of intrinsic stringy origin. We study its phenomenological
characteristics and the consequences in the case that $U(1)_{Z^\prime}$
remains unbroken down to low energies. We suggest that observation
in forthcoming colliders of a $Z^\prime$, with universal couplings
for the two light generations but different couplings for the
heavy generation may provide evidence for the $Z_2\times Z_2$
orbifold which underlies the free fermionic models.

\end{abstract}
\smallskip}
\end{titlepage}

\setcounter{footnote}{0}

\def\at{ }
\def\beq{\begin{equation}}
\def\eeq{\end{equation}}
\def\beqn{\begin{eqnarray}}
\def\eeqn{\end{eqnarray}}
\def\no{\noindent }
\def\nolabel{\nonumber }

\def\NA{non--Abelian }

\def\gsim{{\buildrel >\over \sim}}
\def\lsim{{\buildrel <\over \sim}}

\def\ie{i.e., }
\def\eg{{\it e.g.}\ }
\def\eq#1{eq.\ (\ref{#1})}

\def\lt{<}

\def\slash#1{#1\hskip-6pt/\hskip6pt}
\def\slk{\slash{k}}

\def\dag{\dagger}
\def\qandq{\quad {\rm and} \quad} 
\def\qand{\quad {\rm and} } 
\def\andq{ {\rm and} \quad } 
\def\qwithq{\quad {\rm with} \quad} 
\def\qwith{ \quad {\rm with} } 
\def\withq{ {\rm with} \quad} 

\def\fhalf{\frac{1}{2}}
\def\fsqrt{\frac{1}{\sqrt{2}}}
\def\half{{\textstyle{1\over 2}}}
\def\third{{\textstyle {1\over3}}}
\def\quarter{{\textstyle {1\over4}}}
\def\sixth{{\textstyle {1\over6}}}
\def\m{$\phantom{-}$}
\def\j{$-$}
\def\ps{{\tt +}}
\def\pps{\phantom{+}}

\def\zz{$Z_2\times Z_2$ }

\def\Tr{{\rm Tr}\, }
\def\tr{{\rm tr}\, }

\def\MP{M_{P}}
\def\GeV{\,{\rm GeV}}
\def\TeV{\,{\rm TeV}}

\def\lam#1{\lambda_{#1}}
\def\non{\nonumber}
\def\smgg{ $SU(3)_C\times SU(2)_L\times U(1)_Y$ }
\def\smggb{ $SU(3)_C\times SU(2)_L\times U(1)_Y$}
\def\SM{Standard--Model }
\def\SUSY{supersymmetry }
\def\SSSM{supersymmetric standard model}
\def\MSSM{minimal supersymmetric standard model}
\def\MSSSM{MS$_{str}$SM }
\def\MSSSMc{MS$_{str}$SM, }
\def\obs{{\rm observable}}
\def\sig{{\rm singlets}}
\def\hid{{\rm hidden}}
\def\MS{M_{str}}
\def\Ms{$M_{str}$}
\def\MP{M_{P}}

\def\vev#1{\langle #1\rangle}
\def\mvev#1{|\langle #1\rangle|^2}

\def\UA{U(1)_{\rm A}}
\def\QA{Q^{(\rm A)}}
\def\mssm{SU(3)_C\times SU(2)_L\times U(1)_Y} 

\def\KM{Ka\v c--Moody }

\def\y{\,{\rm y}}
\def\l{\langle}
\def\r{\rangle}
\def\o#1{\frac{1}{#1}}

\def\zi{z_{\infty}}

\def\hb#1{\bar{h}_{#1}}
\def\Htw{{\tilde H}}
\def\chibar{{\overline{\chi}}}
\def\qbar{{\overline{q}}}
\def\ibar{{\overline{\imath}}}
\def\jbar{{\overline{\jmath}}}
\def\Hbar{{\overline{H}}}
\def\Qbar{{\overline{Q}}}
\def\abar{{\overline{a}}}
\def\alphabar{{\overline{\alpha}}}
\def\betabar{{\overline{\beta}}}
\def\tautwo{{ \tau_2 }}
\def\thetatwo{{ \vartheta_2 }}
\def\thetathree{{ \vartheta_3 }}
\def\thetafour{{ \vartheta_4 }}
\def\ttwo{{\vartheta_2}}
\def\tthree{{\vartheta_3}}
\def\tfour{{\vartheta_4}}
\def\ti{{\vartheta_i}}
\def\tj{{\vartheta_j}}
\def\tk{{\vartheta_k}}
\def\calF{{\cal F}}
\def\smallmatrix#1#2#3#4{{ {{#1}~{#2}\choose{#3}~{#4}} }}
\def\ab{{\alpha\beta}}
\def\Minv{{ (M^{-1}_\ab)_{ij} }}
\def\ii{{(i)}}
\def\V{{\bf V}}
\def\N{{\bf N}}

\def\b{{\bf b}}
\def\S{{\bf S}}
\def\X{{\bf X}}
\def\I{{\bf I}}
\def\bone{{\mathbf 1}}
\def\bo{{\mathbf 0}}
\def\bs{{\mathbf S}}
\def\mS{{\mathbf S}}
\def\bS{{\mathbf S}}
\def\bb{{\mathbf b}}
\def\mb{{\mathbf b}}
\def\mX{{\mathbf X}}
\def\bX{{\mathbf X}}
\def\mI{{\mathbf I}}
\def\bI{{\mathbf I}}
\def\balpha{{\mathbf \alpha}}
\def\bbeta{{\mathbf \beta}}
\def\bgamma{{\mathbf \gamma}}
\def\bxi{{\mathbf \xi}}
\def\malpha{{\mathbf \alpha}}
\def\mbeta{{\mathbf \beta}}
\def\mgamma{{\mathbf \gamma}}
\def\mzeta{{\mathbf \zeta}}
\def\mxi{{\mathbf \xi}}
\def\bphi{\overline{\Phi}}

\def\eps{\epsilon}

\def\t#1#2{{ \Theta\left\lbrack \matrix{ {#1}\cr {#2}\cr }\right\rbrack }}
\def\C#1#2{{ C\left\lbrack \matrix{ {#1}\cr {#2}\cr }\right\rbrack }}
\def\tp#1#2{{ \Theta'\left\lbrack \matrix{ {#1}\cr {#2}\cr }\right\rbrack }}
\def\tpp#1#2{{ \Theta''\left\lbrack \matrix{ {#1}\cr {#2}\cr }\right\rbrack }}
\def\l{\langle}
\def\r{\rangle}

\def\op#1{$\Phi_{#1}$}
\def\opp#1{$\Phi^{'}_{#1}$}
\def\opb#1{$\overline{\Phi}_{#1}$}
\def\opbp#1{$\overline{\Phi}^{'}_{#1}$}
\def\oppb#1{$\overline{\Phi}^{'}_{#1}$}
\def\oppx#1{$\Phi^{(')}_{#1}$}
\def\opbpx#1{$\overline{\Phi}^{(')}_{#1}$}

\def\oh#1{$h_{#1}$}
\def\ohb#1{${\bar{h}}_{#1}$}
\def\ohp#1{$h^{'}_{#1}$}

\def\oQ#1{$Q_{#1}$}
\def\odc#1{$d^{c}_{#1}$}
\def\ouc#1{$u^{c}_{#1}$}

\def\oL#1{$L_{#1}$}
\def\oec#1{$e^{c}_{#1}$}
\def\oNc#1{$N^{c}_{#1}$}

\def\oH#1{$H_{#1}$}
\def\oV#1{$V_{#1}$}
\def\oHs#1{$H^{s}_{#1}$}
\def\oVs#1{$V^{s}_{#1}$}

\def\p#1{{\Phi_{#1}}}
\def\pp#1{{\Phi^{'}_{#1}}}
\def\pb#1{{{\overline{\Phi}}_{#1}}}
\def\pbp#1{{{\overline{\Phi}}^{'}_{#1}}}
\def\ppb#1{{{\overline{\Phi}}^{'}_{#1}}}
\def\ppx#1{{\Phi^{(')}_{#1}}}
\def\pbpx#1{{\overline{\Phi}^{(')}_{#1}}}

\def\h#1{h_{#1}}
\def\hb#1{{\bar{h}}_{#1}}
\def\hp#1{h^{'}_{#1}}

\def\Q#1{Q_{#1}}
\def\dc#1{d^{c}_{#1}}
\def\uc#1{u^{c}_{#1}}

\def\L#1{L_{#1}}
\def\ec#1{e^{c}_{#1}}
\def\Nc#1{N^{c}_{#1}}

\def\H#1{H_{#1}}
\def\V#1{V_{#1}}
\def\Hs#1{{H^{s}_{#1}}}
\def\Vs#1{{V^{s}_{#1}}}

\def\fdtv{FD2V }
\def\fdtp{FD2$^{'}$ }
\def\fdtpv{FD2$^{'}$v }

\def\FD2pv{FD2$^{'}$V }
\def\FD2p{FD2$^{'}$ }


\def\inbar{\,\vrule height1.5ex width.4pt depth0pt}

\def\IC{\relax\hbox{$\inbar\kern-.3em{\rm C}$}}
\def\IQ{\relax\hbox{$\inbar\kern-.3em{\rm Q}$}}
\def\IR{\relax{\rm I\kern-.18em R}}
 \font\cmss=cmss10 \font\cmsss=cmss10 at 7pt
 \font\cmsst=cmss10 at 9pt
 \font\cmssn=cmss9

\def\IZ{\relax\ifmmode\mathchoice
 {\hbox{\cmss Z\kern-.4em Z}}{\hbox{\cmss Z\kern-.4em Z}}
 {\lower.9pt\hbox{\cmsss Z\kern-.4em Z}}
 {\lower1.2pt\hbox{\cmsss Z\kern-.4em Z}}\else{\cmss Z\kern-.4em Z}\fi}

\def\AEF{A.E. Faraggi}
\def\AP#1#2#3{{\it Ann.\ Phys.}\/ {\bf#1} (#2) #3}
\def\EPJC#1#2#3{{\it The Euro.\ Phys.\ Jour.\/} {\bf C#1} (#2) #3}
\def\NPB#1#2#3{{\it Nucl.\ Phys.}\/ {\bf B#1} (#2) #3}
\def\NPBPS#1#2#3{{\it Nucl.\ Phys.}\/ {{\bf B} (Proc. Suppl.) {\bf #1}} (#2) 
 #3}
\def\PLB#1#2#3{{\it Phys.\ Lett.}\/ {\bf B#1} (#2) #3}
\def\PRD#1#2#3{{\it Phys.\ Rev.}\/ {\bf D#1} (#2) #3}
\def\PRL#1#2#3{{\it Phys.\ Rev.\ Lett.}\/ {\bf #1} (#2) #3}
\def\PRT#1#2#3{{\it Phys.\ Rep.}\/ {\bf#1} (#2) #3}
\def\PTP#1#2#3{{\it Prog.\ Theo.\ Phys.}\/ {\bf#1} (#2) #3}
\def\MODA#1#2#3{{\it Mod.\ Phys.\ Lett.}\/ {\bf A#1} (#2) #3}
\def\IJMP#1#2#3{{\it Int.\ J.\ Mod.\ Phys.}\/ {\bf A#1} (#2) #3}
\def\nuvc#1#2#3{{\it Nuovo Cimento}\/ {\bf #1A} (#2) #3}
\def\RPP#1#2#3{{\it Rept.\ Prog.\ Phys.}\/ {\bf #1} (#2) #3}
\def\etal{{\it et al.\/}}

\hyphenation{su-per-sym-met-ric non-su-per-sym-met-ric}
\hyphenation{space-time-super-sym-met-ric}
\hyphenation{mod-u-lar mod-u-lar--in-var-i-ant}


\section{Introduction} 

The goal of superstring phenomenology at its present stage
is to develop the tools and methodology to connect between
string theory and experimental data. It is clear that 
to understand the mechanism which selects the string vacuum
a non--perturbative formulation is needed. However, it is rather
likely that detailed confrontation with the experimental data
will have to rely on perturbative means. For this purpose
the tools to construct realistic string models and
the methodology to extract their phenomenological 
implications must be further developed. The first
mandatory task of superstring phenomenology is
to produce string solutions which are as realistic
as possible with present day technology. The goal
in this regard is to construct string models
that aim to reproduce the phenomenological
data provided by the Standard Model spectrum. 
Moreover, with the lack of substantial experimental
evidence for any extension of the Standard Model,
the most desired solution would be one that reproduces
solely the Standard Model. Once the first goal is
achieved the subsequent goal is to extract possible 
experimental signatures, beyond the Standard Model, 
which may provide further evidence for specific
string models, in particular, and for string theory, 
in general. In practice, of course, it makes sense
to try to extract the experimental consequences
of the theory at every stage of its development. 
This, for example, was the drive behind much of the
superstring inspired activity \cite{hewett} that followed the seminal
Candelas \etal~paper \cite{candelas}. 

Pursuing the minimalist approach, and taking the Standard Model
data as the guide toward understanding the basic building blocks
of nature, one is compelled to assess that grand unification
structures are relevant in nature. Proton decay constraints
then imply the big desert scenario, and gravity becomes
important only near the Planck scale. In this eventuality
the perturbative heterotic string, which naturally accommodates
the grand unification structures with chiral matter, 
is the relevant framework. 

Over the past decade the free fermionic formulation \cite{fff} of the
heterotic string has been utilized to derive the most realistic
string models to date \cite{real,fny,nahe}.
A large number of three generation models
have been constructed, which differ in their detailed
phenomenological characteristics. All these models
share an underlying $Z_2\times Z_2$ orbifold structure,
which naturally gives rise to three generation models
with the standard $SO(10)$ embedding of the Standard
Model spectrum \cite{nahe,z2}. In this respect the phenomenological
success of the free fermionic models can be regarded 
as indicating the relevance of the $Z_2\times Z_2$
orbifold structure in nature. Furthermore, recently, 
and for the first time since the advent of superstring
phenomenology, it was demonstrated that free fermionic
models also produce Minimal Standard Heterotic String Models
(MSHSM) \cite{cfn1,cfn2,cfn3,cfn4}.
In such models the low energy spectrum, which is charged
under the Standard Model gauge group, consists solely
of the spectrum of the Minimal Supersymmetric Standard Model.
It should be emphasized that it is not suggested that one
of the free fermionic models that has been constructed
to date is the correct string vacuum. Indeed, such a claim
would at least require a derivation of the detailed fermion mass
spectrum, as well as an understanding of
the dynamics which break supersymmetry.
However, a plausible interpretation of the phenomenological
success of the free fermionic models is that the true string
vacuum is in the neighborhood of these models. This interpretation
would then single out, for example, the $Z_2\times Z_2$ orbifold
as the relevant string compactification. The general lesson
in this respect is to extract the string structures which are
relevant for the Standard Model phenomenological data. 

Subsequent to achieving the mandatory task of demonstrating the
phenomenological viability of a particular class of heterotic
string compactification, trying to extract possible experimental
signatures beyond the Standard Model becomes more compelling.
Various such possible experimental signatures, inspired from string theory,
have been discussed in the past. Among them: additional gauge bosons, exotic
matter, and specific patterns of supersymmetry breaking.
Our aim here is to use the same tools, that have been used
to derive realistic string models, to try to extract the
experimental signatures beyond the Standard Model. Such
experimental signatures then have the advantage of being
``derived'' rather than ``inspired'' from string theory.
In this paper, we focus on the possibility of additional 
gauge bosons, beyond the Standard Model. The possibility of
such additional gauge structure has of course been discussed
extensively in the past, mainly in the context of additional
symmetries which arise in $SO(10)$ and $E_6$ grand 
unification \cite{hewett,leike}.
However, as our analysis demonstrates the most likely
additional gauge bosons to arise from realistic string 
models are not of this origin. Therefore, of particular
interest in our discussion will be the additional gauge bosons
which are of particular string origin, {\it i.e.}, those
that do not arise in Grand Unified theories.

\section{Free fermionic phenomenology}

The analysis of the free fermionic models is conducted in two
steps. In the first step the free fermionic model building
rules are used to construct a consistent three generation string
vacuum. Subsequently one extracts the full massless spectrum
as well as the cubic level and higher order non--renormalizable
terms in the superpotential. At this stage the tools used are
perturbative heterotic string theory techniques.
The superstring derived three generation models contain
numerous massless vector--like states, some of which carry
fractional electric charge. The models typically also contain
a number of additional $U(1)$ symmetries in the observable 
sector, plus a hidden gauge group which is a subgroup of
the original hidden $E_8$ of the heterotic string. One
additionally finds that many of these three generation models
contain an anomalous $U(1)_A$ symmetry, which generates a 
Fayet--Iliopoulos term, 
\beqn
\eps\equiv\frac{g^2_s M_P^2}{192\pi^2}\Tr Q^{(A)},
\label{fit}
\eeqn
where 
$\Tr Q^{(A)}\ne 0\,\, ,$ is the trace of the $U(1)_A$ charge 
over all the massless fields. The Fayet--Iliopoulos term
breaks supersymmetry near the Planck scale, and destabilizes
the string vacuum. Supersymmetry is restored and the vacuum is
stabilized if there exists a direction in the scalar potential 
$\phi=\sum_i\alpha_i\phi_i$ which is $F$--flat and also
$D$--flat with respect to the non--anomalous gauge symmetries
and in which $\sum_i Q_i^A\vert\alpha_i\vert^2$ and $\eps$
are of opposite sign.
If such a direction exists it will acquire a vacuum expectation
value (VEV)
cancelling the anomalous $D$--term, restoring supersymmetry and
stabilizing the string vacuum. Since the fields
corresponding to such a flat direction typically
also carry charges for the non--anomalous $D$--terms, a
non--trivial set of constraints,
\beqn
\vev{D_{A}} &=& \sum_m Q^{(A)}_m |\vev{\varphi_{m}}|^2 +
\eps  = 0\,\, , \label{daf}\\
\vev{D_i} &=& \sum_m Q^{(i)}_m |\vev{\varphi_{m}}|^2 = 0\,\, .
\label{dana}
\eeqn
on the possible choices of VEVs is imposed. These scalar
VEVs will in general break some, or all, of the additional
symmetries spontaneously. 

Additionally one must insure that
the supersymmetric vacuum is also $F$--flat. 
Each superfield $\Phi_{m}$ (containing a scalar field $\varphi_{m}$
and chiral spin--$\half$ superpartner $\psi_m$) that appears
in the superpotential imposes further constraints on the scalar VEVs. 
$F$--flatness will be broken (thereby destroying spacetime supersymmetry) at 
the scale of the VEVs unless,
\beq
\vev{F_{m}} \equiv \vev{\frac{\partial W}
{\partial \Phi_{m}}} = 0; \,\, \vev{W}  =0,
\label{ff}
\eeq
where $W$ is the superpotential which contains cubic level and
higher order non--renormalizable terms. 
The higher order terms have the generic
form $$<\Phi_1^f\Phi_2^f\Phi_3^b\cdots \Phi_N^b>.$$ Some of the
fields appearing in the non--renormalizable terms will
in general acquire a non--vanishing VEV by the anomalous
$U(1)$ cancellation mechanism. Thus, in this process some of the 
non--renormalizable terms induce effective renormalizable
operators in the effective low energy field theory 
wherein either all fields or all fields but one are replaced with VEVs.  
One must insure that such terms do not violate supersymmetry
at an unacceptable level. In practice, however, the studies
performed to date have been restricted to the case in which
supersymmetry remains unbroken to all orders of 
non--renormalizable terms.

Thus, the second stage of the string model building
analysis is conducted by analyzing the $F$-- and $D$--flat
directions. An important advance of the last few years
has been the development of systematic techniques for
the analysis of exact $F$-- and $D$--flat directions \cite{langacker}.
Additionally one may impose other phenomenological 
constraints on the scalar VEVs. For example, in demonstrating
the existence of a free fermionic MSHSM, we required 
that a set of fields which induces the decoupling
of all non--MSSM states, acquire a non--vanishing
VEV along the $F$-- and $D$--flat directions. 
The string model building analysis outlined above
can be regarded as aiming to achieve the first
goal of superstring phenomenology. Namely, to
reproduce the data provided by the Standard Model.

\section{Additional gauge symmetries in free fermionic models}

In this section we 
discuss the different classes of additional gauge symmetries
that are obtained in the free fermionic models prior to the analysis of
flat directions. 
For a given three generation string model, and
a specific flat direction, the resulting string vacuum
may give rise to additional matter and gauge 
bosons, which are beyond the Minimal Supersymmetric
Standard Model. For example, the hidden sector
may give rise to massless matter states, which
are not charged with respect to the Standard Model
gauge group, and which interact with the Standard Model
states only via horizontal $U(1)$ symmetries. 
Such hidden matter states may have interesting 
cosmological implications and may serve as dark matter
candidates. Similarly, for a given flat direction
the string vacuum may contain a combination of the
horizontal $U(1)$ symmetries, which remains
unbroken. It is this type of unbroken $U(1)$ symmetry
that we aim to study in this paper. Thus, for a given
flat direction the first task is to extract the
combinations of the $U(1)$ symmetries which remain
unbroken. Some of the resulting combinations
may be entirely hidden. Namely, the Standard Model
states will not be charged under them. Such
hidden combinations may therefore be less interesting
from an experimental point of view. However, there
may also exist unbroken combinations of the
horizontal $U(1)$ symmetries, under which
the Standard Model states are charged. It is
precisely this type of unbroken $U(1)$ symmetries
that are of enormous experimental and phenomenological
interest. Furthermore, in a given string model
the charges of the Standard Model states under
such an unbroken $U(1)$ symmetry are completely specified.
Consequently, the phenomenology of the additional $Z^\prime$
in a given string model is specified, up to some
educated assumptions on the scale of $Z^\prime$ breaking
and the strength of its coupling. Both of these assumptions
will of course eventually be relaxed. 

The free fermionic models are constructed by specifying a set
of boundary conditions basis vectors and the one--loop
GSO projection coefficients \cite{fff}.
The basis vectors, $\mb_k$, span a finite  
additive group $\Xi=\sum_k{{n_k}{\mb_k}}$
where $n_k=0,\cdots,{{N_{z_k}}-1}$, with 
$N_{z_k}$ the smallest positive integer such that
$N_{z_k} \mb_k = \vec{0}$ (mod 2). 
The physical massless states in the Hilbert space of a given sector
$\balpha\in{\Xi}$, are obtained by acting on the vacuum with 
bosonic and fermionic operators and by
applying the generalized GSO projections. The $U(1)$
charges, $Q(f)$, with respect to the unbroken Cartan generators of the four 
dimensional gauge group, which are in one 
to one correspondence with the $U(1)$
currents ${f^*}f$ for each complex fermion f, are given by:
\beqn
{Q(f) = {1\over 2}\alpha(f) + F(f)},
\label{u1charges}
\eeqn
where $\alpha(f)$ is the boundary condition of the world--sheet fermion $f$
in the sector $\alpha$, and 
$F_\alpha(f)$ is a fermion number operator counting each mode of 
$f$ once (and if $f$ is complex, $f^*$ minus once). 
For periodic fermions,
$\alpha(f)=1$, the vacuum is a spinor in order to represent the Clifford
algebra of the corresponding zero modes. 
For each periodic complex fermion $f$
there are two degenerate vacua ${\vert +\rangle},{\vert -\rangle}$ , 
annihilated by the zero modes $f_0$ and
${{f_0}^*}$ and with fermion numbers  $F(f)=0,-1$, respectively. 

The four dimensional gauge group in the three generation
free fermionic models arises as follows. The models can 
in general be regarded as constructed in two stages.
The first stage consists of the NAHE set of boundary conditions basis
vectors, which is a set of five boundary condition basis vectors, 
$\{{\bf1},\bS,\mb_1,\mb_2,\mb_3\}$ \cite{nahe}. The NAHE set is a common
set in the three generation models that we discuss here.
The gauge group after imposing the GSO projections induced
by the NAHE set basis vectors is $$SO(10)\times SO(6)^3\times E_8$$
with $N=1$ supersymmetry. The space--time vector bosons that generate
the gauge group arise from the Neveu--Schwarz sector and
from the sector ${\bf1}+\mb_1+\mb_2+\mb_3$. The Neveu--Schwarz sector
produces the generators of $SO(10)\times SO(6)^3\times SO(16)$.
The sector $\mzeta\equiv{\bf1}+\mb_1+\mb_2+\mb_3$ produces the spinorial of 
$\mathbf 128$
of $SO(16)$ and completes the hidden gauge group to $E_8$.
At the level of the NAHE set the sectors $\mb_1$, $\mb_2$ and $\mb_3$
produce 48 multiplets, 16 from each, in the $\mathbf 16$ 
representation of $SO(10)$.

We remark that in order to understand the origin of
the various additional $U(1)$ symmetries
that may appear in free fermionic models it is often useful to consider
a version of the NAHE set, which is extended by adding the basis
vector $X$ with periodic boundary conditions for the right--moving
complex fermions $\{{\bar\psi}^{1,\cdots,5},{\bar\eta}^{1,2,3}\}$
\cite{nahe,z2}.
With this additional boundary basis vector the four dimensional
gauge symmetry is extended to $$E_6\times U(1)^2\times SO(4)^3\times
E_8.$$ One can regard this set as starting with a toroidally
compactified model generated by the set of basis vectors $\{{\bf1},\bS,\bX,
\mzeta\}$. The right--moving gauge group with this set is
$SO(12)\times E_8\times E_8$, with $N=4$ space--time supersymmetry.
The basis vectors $\mb_1$ and $\mb_2$ are then used to
break $N=4\rightarrow N=1$ supersymmetry, and to reduce the gauge
symmetry to $E_6\times U(1)^2\times SO(4)^3\times E_8$.  
The $U(1)$ combination produced by the world--sheet currents
$\bar\eta^1\bar\eta^{1^*}+\bar\eta^2\bar\eta^{2^*}+\bar\eta^3\bar\eta^{3^*}$
becomes the $U(1)$ symmetry in the decomposition of $E_6$ under
$SO(10)\times U(1)$. The realistic free fermionic models 
can be regarded as starting with this set, but changing the
sign of the GSO phase $c({\mzeta\atop{\mX}})$. With this GSO phase
change the ${\mathbf 16}+{\overline{\mathbf 16}}$ generators in the adjoint
of $E_6$ are projected out. The right--moving gauge group in this
case becomes $SO(10)\times U(1)_A\times U(1)^2\times SO(4)^3\times
E_8$, with $U(1)_A$ being anomalous \cite{cf1}. 

The second stage of the free fermionic
basis construction consists of adding to the 
NAHE set three (or four) additional boundary condition basis vectors. 
These additional basis vectors reduce the number of generations
to three chiral generation, one from each of the basis vectors $\mb_1$,
$\mb_2$ and $\mb_3$, and simultaneously break the four dimensional
gauge group. The $SO(10)$ is broken to one of its subgroups
$SU(5)\times U(1)$, $SO(6)\times SO(4)$ or $SU(3)\times SU(2)\times U(1)^2$.
Similarly, the hidden $E_8$ symmetry is broken to one of its
subgroups by the basis vectors which extend the NAHE set.
This hidden $E_8$ subgroup may, or may not, contain $U(1)$ factors
which are not enhanced to a non--Abelian symmetry. On the other
hand, the flavor $SO(6)^3$ symmetries in the NAHE--based models
are always broken to flavor $U(1)$ symmetries, as the breaking
of these symmetries is correlated with the number of chiral
generations. Three such $U(1)_j$ symmetries are always obtained
in the NAHE based free fermionic models, from the subgroup
of the observable $E_8$, which is orthogonal to $SO(10)$.
These are produced by the world--sheet currents ${\bar\eta}{\bar\eta}^*$
($j=1,2,3$), which are part of the Cartan sub--algebra of the
observable $E_8$. Additional unbroken $U(1)$ symmetries, denoted
typically by $U(1)_j$ ($j=4,5,...$), arise by pairing two real
fermions from the sets $\{{\bar y}^{3,\cdots,6}\}$,
$\{{\bar y}^{1,2},{\bar\omega}^{5,6}\}$ and
$\{{\bar\omega}^{1,\cdots,4}\}$. The final observable gauge
group depends on the number of such pairings. 

Our interest in this paper is in additional gauge bosons
that arise from the horizontal flavor symmetries. That is,
in additional vector bosons which arise from combinations
of world--sheet $U(1)$ currents of the Cartan subalgebra. 
The generators of such additional $U(1)$'s all arise 
from the Neveu--Schwarz sector. Before proceeding, however,
we briefly discuss possible non--Abelian extensions
of the Standard Model in these models, and postpone
detailed analysis on these possibilities to future work.
It is already clear that from the unbroken subgroup of
$SO(10)$, we can obtain the traditional left--right 
symmetric extensions of the Standard Model. These
originate from the $SO(6)\times SO(4)$ type models
or from left--right symmetric models in which
$SO(10)$ is broken to $SU(3)\times U(1)\times SO(4)$
at the string level. 

Additional sources of possible
non--Abelian enhancement may arise from combinations
of the boundary conditions basis vectors which
extend the NAHE set. In some of the three generation
model one finds combinations of the additional basis vectors
\beqn
Y=n_\alpha\balpha+n_\bbeta\beta+n_\bgamma\gamma,
\label{ydef}
\eeqn
for which $Y_L\cdot Y_L=0$ and $Y_R\cdot Y_R\le8$.
Such a combination may produce additional space--time
vectors bosons, depending on the GSO projections.
In these cases, some combination of the $U(1)$ generators
of the four dimensional Cartan sub--algebra is enhanced
to a non--Abelian gauge symmetry. Often it is
found that this is a combination of the flavor
symmetries, which is family universal, and combines
with the $U(1)_{B-L}$ generator to produce a baryonic,
or leptonic, non--Abelian gauged symmetry
\cite{leptophobic}. Such
symmetries may therefore play an important role in insuring
proton stability, but their phenomenological
viability still needs to be studied. We will not discuss
this type of enhanced symmetries further here and delegate
more detailed studies to future work.
To summarize, in the spirit of the minimalist approach pursued
here, in this paper we are interested in additional gauge bosons
that arise from the unbroken Cartan generators of the four dimensional
gauge group. In particular, we are interested in possible combinations
of the $U(1)$ symmetries, which remain unbroken by a set
of flat directions that cancels the anomalous $U(1)$ $D$--term.

\section{$Z^\prime$s in free fermionic models}

We now turn to discuss the extra $Z^\prime$ symmetries
that appear in free fermionic models. The first objective is to
study the additional $U(1)$ symmetries which appear prior to
the analysis of flat directions.
The subsequent 
objective is to determine which combinations of $U(1)$ symmetries
remain unbroken after the analysis of flat directions, 
and possibly after imposing additional phenomenological
constraints that are required by the Standard Model data.
Such unbroken $U(1)$ combinations then come closer to being
a prediction of the string models. The final goal is of course
to extract which possible $U(1)$ combinations remain unbroken
after the wealth of Standard Model experimental data is satisfied. 
Such an extra $U(1)$ combination then is truly a prediction
of a specific string vacua. However, short of this ambitious
and still unachievable goal, we can already at this stage
extract the general characteristics of $U(1)$ combinations
that may remain unbroken in detailed $F$-- and $D$--flat solutions.

The first type of $Z^\prime$ symmetry
that has been considered in the context of free fermionic models
\cite{zprime} has been the $U(1)$ combination 
\beq
Q_{Z^\prime}={{B-L}\over2}-{2\over3}T_{3_R},
\label{qzprime1}
\eeq
which is embedded in $SO(10)$ and is orthogonal to the
Standard Model weak--hypercharge. The phenomenology of
this class of extra $U(1)$'s, as well as its family--universal
extensions in the context of $E_6$ string inspired 
phenomenology, have been discussed extensively in the past
\cite{hewett,leike}. 
As we discussed above, in the free fermionic models
the additional $U(1)$ symmetry (aside from (\ref{qzprime1})
which is embedded in $E_6$ is given by the 
family universal combination of the horizontal symmetries,
given by 
\beqn
U(1)_{E_6}=U(1)_1+U(1)_2+U(1)_3.
\label{ue6}
\eeqn
However, there are several reasons to argue
that these particular $U(1)$ combinations,
(\ref{qzprime1}) and (\ref{ue6}),
in the free fermionic
models cannot remain unbroken to low energies. In the first
place, one often finds (although not always \cite{cfs})
that the family universal $U(1)$ which is 
embedded in $E_6$ is anomalous and is therefore broken
by the flat direction VEVs. Second, the scale of the breaking
of the $U(1)$ symmetry which is embedded in $SO(10)$ is
associated with the see--saw scale, which is needed to suppress
the left--handed neutrino masses. Thus, the requirement of
sufficiently small neutrino masses implies that this particular
$U(1)$ symmetry cannot remain unbroken to low energies. 

The natural question is then which additional $U(1)$ symmetries,
beyond the weak--hypercharge of the Standard Model, can remain
unbroken to low energies. As discussed in the introduction
the string models under considerations often contain an 
anomalous $U(1)$ symmetry. In those cases most, or all, of the
horizontal $U(1)$ symmetries in the observable sector are
broken by the choices of flat directions. Additionally, 
one has to impose plausible phenomenological constraints,
like the decoupling of exotic fractionally charged states
and quasi--realistic fermion mass spectrum, which may further
result in the breaking of the observable horizontal symmetries.
The choice of flat directions may leave unbroken $U(1)$ symmetries
in the hidden sector, but those are of less interest from an
experimental and phenomenological perspective. We further
remark that in left--right symmetric models, with $SU(3)\times
U(1)\times SU(2)_L\times SU(2)_R$ as the unbroken subgroup of $SO(10)$
at the string scale, one finds models in which all the 
horizontal $U(1)$ symmetries are anomaly free \cite{cfs}.
That is, in these models there is no anomalous $U(1)$ symmetry.
Consequently these semi--realistic string vacua are supersymmetric
and anomaly free without the need for scalar VEVs which break
some of the horizontal $U(1)$ symmetries. However, the phenomenology
of this class of string models has not been studied extensively and
one may expect that imposing plausible phenomenological
constraints will necessitate some Planck scale VEVs. Therefore,
in this paper we focus on string models that do contain
an anomalous $U(1)$ symmetry.

\section{$Z^\prime$ in the FNY model}

As our concrete illustrative 
example of a $Z^\prime$ appearing in a string model
we consider the string derived model of ref.~\cite{fny}. We will refer
to this model as the FNY model. The $F$-- and $D$--flat
directions in this model were studied in detail in 
refs.~\cite{cfn1,cfn2,cfn3,cfn4}. There it was shown that there
exist for this model flat directions which result in the
decoupling of all the massless exotic fractionally charged states
by the scalar VEVs. This is achieved due to the fact that
in this model there exist cubic level superpotential 
terms, in which the exotic fractionally charged states
are coupled to a set of $SO(10)$ singlets. Thus, assigning
non--vanishing VEVs to this set of $SO(10)$ singlets
results in all of the fractionally charged exotic states
receiving mass of the order of the Fayet--Iliopoulos
term. It was further shown that for these flat
directions all the additional states beyond the spectrum of
the Minimal Supersymmetric Standard Model receive masses
from up to quintic order terms in the superpotential.
Therefore, in this model all the states that are beyond the MSSM
and which are charged with respect to the Standard Model gauge group 
decouple from the massless spectrum at or
slightly below the string scale. This string model therefore
provides the first known example of a Minimal Standard 
Heterotic--String Model (MSHSM). It should be emphasized
that this does not indicate that the FNY string model
is the correct string vacuum, nor is it our intention
to claim that the FNY model passes all of the phenomenological
constraints imposed by the Standard Model data. However, what
we think is a reasonable lesson to extract is that
the success of producing a MSHSM, as well as the
other unique phenomenological characteristics of the
free fermionic models, like the standard $SO(10)$
embedding of the weak--hypercharge, may be taken as suggesting
that the correct string vacuum may indeed exist in the vicinity
of the free fermionic point in the string moduli space.
The details of the FNY string model, its massless spectrum, 
and superpotential
terms up to sixth order are given in ref.~\cite{fny,cfn1,cfn2}.
Here, for completeness, we only discuss the features of the model
which are relevant for our discussion, and give in Table 1 the 
relevant states and charges in the effective low
energy field theory.

Prior to the analysis of flat directions the observable gauge group
of the FNY model is: $SU(3)_C\times SU(2)_L
\times U(1)_C\times U(1)_L\times U(1)^6$
\footnote{$U(1)_C={3\over2}U(1)_{B-L}$; 
$U(1)_L=2U(1)_{T_{3_R}}$.}, and
the hidden gauge group is: $SO(4)\times SU(2)\times U(1)^4$.
The Standard Model weak--hypercharge is given by
$U(1)_Y= \frac{1}{3} U(1)_C+\frac{1}{2}U(1)_L$.
The sectors $\mb_1$, $\mb_2$ and $\mb_3$ produce the three light generations.
Electroweak Higgs doublets $\{h_{1,2,3},{\bar h}_{1,2,3}\}$
arise from the Neveu--Schwarz sector, and $H_{34}$, $H_{41}$ from
the sectors $\mb_3+\alpha\pm\beta$ and $\mb_1+\mb_2+\mb_4+\balpha\pm\bbeta$.

Prior to rotating the anomaly into a single $\UA$, 
six of the FNY model's twelve $U(1)$ symmetries are anomalous:
Tr${\, U_1=-24}$, Tr${\, U_2=-30}$, Tr${\, U_3=18}$,
Tr${\, U_5=6}$, Tr${\, U_6=6}$ and  Tr${\, U_8=12}$.
Thus, the total anomaly can be rotated into a single 
$U(1)_{\rm A}$ defined by 
\beq
U_A\equiv -4U_1-5U_2+3U_3+U_5+U_6+2U_8.
\label{anomau1infny}
\eeq
The five orthogonal linear combinations,
\beqn
U^{'}_1 &=& \hbox to 3.0truecm{$2 U_1  - U_2 + U_3$\,\, ;\hfill}\quad 
U^{'}_2= -U_1 + 5 U_2 + 7 U_3\,\, ;\nolabel\\
U^{'}_3 &=& \hbox to 3.0truecm{$U_5 - U_6$\,\, ;\hfill}\quad 
U^{'}_4= U_5 + U_6 - U_8\,\, ; \label{nonau1}\\
U^{'}_5 &=& 12 U_1 + 15 U_2 - 9 U_3 + 25 U_5 + 25 U_6 + 50 U_8\,\, ,
\nolabel
\eeqn
are all traceless.

A particular flat solution in the FNY model is given by the set
of fields
\beq
\{\Phi_{12},\Phi_{23},{\bar\Phi}_{56},\Phi_4,{\Phi}_4^\prime,
{\bar\Phi}_4,{\bar\Phi}_4^\prime,H_{31},H_{38},H_{23},V_{40},
H_{28},V_{37}\}.
\label{setofvevs}
\eeq
As discussed in ref.~\cite{cfn3,cfn4} with this set of VEVs all of the
exotic states beyond the MSSM receive heavy mass from cubic or quintic
order terms in the superpotential. 

Detailed investigation of the fermion mass texture which is
generated by the $F$-- and $D$--flat solutions has been performed
in ref.~\cite{cfn3}. The analysis was performed for flat
directions which utilize only non--Abelian singlet VEVs. The
solution in Eq.~(\ref{setofvevs}) also contains non--Abelian
fields, and was shown to be flat to all orders in ref.~\cite{cfn4}.
Quick examination of the non--renormalizable terms suggests that
the fermion mass textures generated by this flat direction
are similar to those found in ref.~\cite{cfn2}. We then have
that the light Higgs representations consist of ${\bar h}_1$
and a combination of $h_1$ and $h_3$. One then finds that
the leading mass terms are $Q_1u_1^c{\bar h}_1$ and
$Q_3d_3^ch_3$. These mass textures are therefore not phenomenologically
viable as the left--handed component of the top and bottom quarks
live in different multiplets. A plausible solution is to find
a flat direction for which $h_3$ is not part of the surviving
light Higgs combination. In which case a mass term for the bottom
quark can appear, for example, from the quartic term
$Q_1d_1^cH_{41}H_{21}^s$.
For the purpose of our discussion here we make the assumption that
the sector $\mb_1$ produces the heavy generation states and $\mb_{2,3}$
produce the two light generations. A more detailed study of the
phenomenology of the non--Abelian flat directions
will be reported in ref.~\cite{cfn5}. 

We next turn to discuss whether any combination, and which, of the
$U(1)$ symmetries of the FNY model remains unbroken by the choice
of VEVs in Eq.~(\ref{setofvevs}). Subsequently, we will examine
the phenomenological characteristics of the unbroken combinations.
The first observation is that the family universal $U(1)_{Z^\prime}$,
which is embedded in $SO(10)$ is broken by this choice of VEVs. 
Similarly, the family universal $U(1)$ combination which is embedded
in $E_6$ is broken at the string scale. As we discussed above, our
general expectation is that in fact these particular $U(1)$ 
symmetries cannot remain unbroken to low energies.
The set of VEVs in Eq.~(\ref{setofvevs}) leaves two $U(1)$
combination unbroken at the string scale. The first is given by
the combination
\beq
U(1)=3U(1)_7+U_h~,
\label{firstu1}
\eeq
while the second unbroken combination is given by
\beq
U(1)_{Z^\prime}=5U(1)_3^\prime+U(1)_7-3U_h~.
\label{secondu1}
\eeq
The $U(1)$ generators appearing in the first combination are from
the Cartan sub--algebra of the hidden $E_8$. Therefore,
the three Standard Model generations from the sectors
$\mb_1$, $\mb_2$ and $\mb_3$ are not charged with respect to
this $U(1)$ combination and it is consequently not
of interest from the point of view of low energy experiments.
In the second unbroken combination $U(1)_3^\prime$ appears
and consequently the Standard Model states are charged
under this unbroken $U(1)_{Z^\prime}$ symmetry. 

\section{Phenomenological characteristics}

As we illustrated in the previous section, the $F$-- and $D$--flat 
solution Eq.~(\ref{setofvevs}) leaves the $U(1)_{Z^\prime}$ combination,
Eq.~(\ref{secondu1}), unbroken at the string scale.
Several issues are important to consider in regard to the
possible low energy phenomenological implications.
Furthermore, many of the issues which are crucial
for fully extracting the phenomenological consequences,
like the fermion identification, are still not under
complete control. Consequently, prior to embarking
on a detailed phenomenological analysis, we have to
try to isolate those characteristics which are independent
of the details about which we are ignorant at this stage.
If such an extraction is possible then the discussion becomes
more substantial. This is the price we have to pay for
trying to extract phenomenological consequences from
a theory whose natural scale is vastly separated from
the experiments' natural scale. Similarly, to this level
we have found a $U(1)$ combination which remains
unbroken at the string scale. It is quite plausible that
supersymmetry breaking requires the existence of an intermediate
energy scale. Of course, one can devise various scenarios,
like radiative breaking, by which the $U(1)_{Z^\prime}$
will be broken just at the right scale, namely near
the electroweak scale. The $U(1)_{Z^\prime}$ breaking
can be generated due to the VEV of one of the remaining light Standard Model
singlets, which are charged under $U(1)_{Z^\prime}$, for example
$\Phi_{56}^\prime$ and ${\bar\Phi}_{56}^\prime$.
But at this stage we regard the
possibility that the $U(1)_{Z^\prime}$ remains unbroken down
to low energies as an assumption and extract the phenomenological
implications from there. We see from Table 1
that the charges of the three generations and the Higgs multiplets
under the $U(1)_{Z^\prime}$ are completely
specified. Then, up to the caveat stated above,
the phenomenological implications are completely fixed.

Several observations are interesting to note.
First from Eq.~(\ref{secondu1}) we see that indeed the unbroken
$U(1)_{Z^\prime}$ is not of $E_6$ or $SO(10)$
origin. Moreover, the unbroken $U(1)$ combination
does not arise from the $U(1)$ generators of the
observable $E_8$, but rather from $U(1)$ symmetries
which arise from the compactified Narain lattice.
Thus, the unbroken $U(1)$ symmetries
that we may expect to arise from string vacua
are not of the GUT type. Furthermore, as the
fermion charges are related to the particular
type of compactification, $U(1)_{Z^\prime}$
experimental data may contain information on
the underlying compactified manifold. 
 
Examining then the $U(1)_{3^\prime}$
charges in Table 1 we see that mass mixing of the $Z^\prime$--gauge
boson with the Standard Model $Z$ would not arise if
the light electroweak doublets are composed only
of doublets from the Neveu--Schwarz sector.
This in fact would be the general case if the
unbroken $U(1)$ is solely a combination of
the Cartan generators arising from the Narain lattice, and
possibly hidden sector generators. That is, if
it does not contain $U(1)$ currents from the observable
$E_8$. In this case, as is seen from Table 1, 
all Neveu--Schwarz electroweak doublets are neutral
with respect to $U(1)_{Z^\prime}$. $Z-{Z^\prime}$
mass mixing could arise if the light electroweak Higgs
doublets contain a state which arises 
from the twisted sectors. In the case of the FNY model
those are $H_{34}$ and $H_{41}$ in Table 1.
In general, the states of this type, arising
from twisted sectors, are charged with respect to
the $U(1)$ currents which arise from the Narain lattice.
However, here it is found that also $H_{34}$ and
$H_{41}$ are neutral under the particular unbroken
$U(1)$ combination given in Eq.~(\ref{secondu1}).
Therefore, here all the electroweak Higgs doublets
are neutral under $U(1)_{Z^\prime}$ and $Z-Z^\prime$
mass mixing cannot arise. 

Possible kinetic mixing, 
arising from one--loop oblique corrections to the
gauge boson propagator, is also highly suppressed.
This follows from our assumption that the sector
$\mb_1$ produces the heavy generation. The heavy generation states
are therefore neutral under $U(1)_{Z^\prime}$, 
and do not contribute to the one--loop corrections.
For the two light generations, arising from
the sectors $\mb_2$ and $\mb_3$, we see from Table 1
that the charges of the states are equal in magnitude
and opposite in sign, and would therefore cancel.
For Standard Model fermions 
the kinetic mixing can therefore only be of the order
of $(\ln (m_1^2 m_2^2)/M_{Z^\prime}^2)\sim(\ln(m_c^2 m_u^2)/M_{Z^\prime}^2)$, 
which is highly suppressed even for $M_{Z^\prime}\sim 500$ GeV.
Gauginos, Higgsinos and the light Higgs cannot contribute to kinetic
mixing because they are all neutral under this particular $U(1)_{Z^\prime}$.

We now turn to the supersymmetric scalar sector. Since under our assumption
the sector $\mb_1$ produces the heavy generation, which is neutral
under $U(1)_{Z^\prime}$, only the two light generation can contribute.
However, up to light fermion mass corrections, and
assuming universality, the two light scalar generations
are degenerate in mass. 
Nonuniversality, could arise due to the $U(1)_{Z^\prime}$ $D$--term
contribution. However, $D_{Z^\prime}$ vanishes if the VEVs
of the two fields which break $U(1)_{Z^\prime}$,
say $\Phi_{56}^\prime$ and ${\bar\Phi}_{56}^\prime$, are equal
in magnitude. Therefore, under this assumption,
the scalar contribution to the
scalar masses is also negligible, and kinetic mixing is highly suppressed
for this particular $Z^\prime$ combination.

We now give a rough estimate of the phenomenological
constraints on $M_{Z^\prime}$. For this purpose we
have to normalize $U(1)_{Z^\prime}$ so that it has the 
correct normalization to produce the correct conformal
dimension, ${\bar h}=1$, for the massless states.
>From Eq.~(\ref{secondu1}) we deduce that the normalization
factor is $N=1/\sqrt{78}$. Estimating the beta function coefficients
from the charges given in Table 1, we obtain $b_{Z^\prime}\approx 2.4$,
where we have taken the spectrum to consist of three 
MSSM generations, excluding the three right--handed neutrinos. 
Taking $\alpha_{\rm GUT}^{-1}\approx25$, and extrapolating from
$M_{\rm GUT}\sim 10^{17}$ GeV to $M_Z$, we obtain
$\alpha_{Z^\prime}^{-1}(M_Z)\approx 40$. As seen from Table 1
the charges of the two light generations, while equal in magnitude
are opposite in charge, and consequently not universal. Therefore,
the strongest constraint is from Flavor Changing Neutral Currents,
which arises here from fermion mixing. To estimate this
constraint we use 
\beqn
\Gamma(K_L^0\rightarrow\mu^+\mu^-)\approx
10^{-8}\Gamma(K^+\rightarrow\mu^+\nu_\mu)\, .
\label{csa} 
\eeqn
Estimating the
tree diagrams we obtain 
\beqn 
Q_{Z^\prime}^4\alpha_{Z^\prime}^2\cos^2\theta_C
\sin^2\theta_C/M_{Z^\prime}^4=10^{-8}\alpha_2^2\sin^2\theta_C/(4M_W^4)\, .
\label{qzp}
\eeqn
With the appropriately normalized charges
for $Q_{Z^\prime}$, we obtain 
$M_{Z^\prime}\approx 25 M_W\approx 2\, {\rm TeV}\,.$
We remark that the additional suppression of the $Z^\prime$ interaction
is obtained because of the $U(1)_{Z^\prime}$
normalization factor that we calculated
above. This reflects the fact that the $U(1)_{Z^\prime}$ combination
contains Cartan generators of the hidden $E_8$ under which the Standard
Model states are not charged. The consequence is that there is
roughly an order of magnitude suppression of the $Q_{Z^\prime}$
charges of the Standard Model states. 

A more stringent constraint arises by considering the mixing in the
$K_0-{\overline K}_0$ system parametrized by the mass difference
$\Delta M_K=3.5\times 10^{-12}\, {\rm MeV}$ \cite{pdg}. 
Treating the $Z^\prime$ as a contact interaction we have 
that $\Delta M_K\sim G_2 M_K f_K^2$, where $f_K\approx1.2m_\pi$
is the kaon decay constant, $M_K\approx 0.5$ GeV is the
kaon mass, and $G_2=(Q_{Z^\prime}^2/M_{Z^\prime}^2)4\pi\alpha_{Z^\prime}
(\cos\theta_C\sin\theta_C)^2$ is the $Z^\prime$ contact
interaction term. We then find that $G_2\le10^{-7}G_F$,
where $G_F$ is the Fermi constant. From this we obtain
$M_{Z^\prime}>30$ TeV.
Considering the corresponding mass difference in
the $B$--meson system, $\Delta M_{B}\approx3.12\cdot10^{-4}{\rm eV}$
\cite{pdg},
imposes only $m_{Z^\prime}\ge500{\rm GeV}$, and is therefore
less restrictive, where we have used the Standard Model
value for $V_{td}$.
We do not estimate
here constraints arising from FCNC in the lepton sector
as the leptonic mixing parameters are not known. 

It is therefore expected that a $Z^\prime$ with non--universal
charges for the two light generations is constrained to be 
above the reach of the LHC. Nevertheless, as we discuss
below, a $Z^\prime$ with universal couplings for the first two
light generations and with different couplings to the
heavy generation may also arise from the free fermionic models
and may in fact be a signature of the $Z_2\times Z_2$ orbifold
which underlies the free fermionic models.   
We note that if a $Z^\prime$ gauge boson of the type that we discussed
above is in the region accessible to future hadron colliders, it will
yield spectacular signatures. Namely, in the case of the 
particular $U(1)_{Z^\prime}$ combination that we examined
here, it will result in enhancement in the production of 
the two light generations, whereas a parallel enhancement
in the production of the heavy generation will not be observed.
Similarly, for this particular $U(1)_{Z^\prime}$ combination,
production of Higgs doublets and gauginos in the $Z^\prime$
channel will not be observed. While it is not our aim to
argue that the particular $U(1)_{Z^\prime}$ combination
examined here is necessarily phenomenologically viable,
what we see is that in a given string model, and for a given
flat direction, the phenomenological consequences and possible
production and decay channels are completely specified and
yield distinctive signatures.

\section{Discussion}

We emphasize that it is not our intent to argue here that the
particular $U(1)_{Z^\prime}$ combination that we examined
is necessarily ``the'' phenomenologically
viable combination that may be seen in future
collider experiments. What we have shown is that in a specific
string model the $U(1)_{Z^\prime}$ combinations which remains unbroken
for specific flat directions are given. Consequently, 
the charges of the Standard Model fermions are specified and,
in the case that the $U(1)_{Z^\prime}$ symmetry remains unbroken
down to low energies, the phenomenological implications are determined.
The $U(1)_{Z^\prime}$ that we examined here provides an illustrative 
example. However, we believe that more general lessons can
be extracted. The first is that we anticipate that the $U(1)$
combinations which remain unbroken after analysis of the
flat directions are not of the type which appear in $SO(10)$
or $E_6$ grand unifying theories. Therefore, it is
anticipated that if a $U(1)$ combination remains unbroken
down to low energies, it contains $U(1)$ factors which
are external to the GUT gauge group.

The second important lesson arises by examining the various
$U(1)$ charges given in Table 1.
We see that a common feature is precisely the flavor
non--universality of the different $U(1)$ combinations.
Thus, we see, for example that for $U_1^\prime$, $U_4^\prime$, and
$U_4$ the charges of the two light generations are universal
and differ from the charges of the heavy generation.
Flat directions which preserve one of these $U(1)$'s 
as a component of an unbroken $U(1)$ symmetry,
may therefore yield a $Z^\prime$ gauge boson which is
less severely restricted by FCNC constraints.
Nevertheless, the distinctive collider signatures
of a $Z^\prime$ arising from any of those $U(1)$ symmetries
will be a non--universality in the production of the different
generations. Thus, for example, $U_4^\prime$ would predict
enhancement in the production of the two light families,
without a corresponding enhancement in production
of the heavy family, whereas $U_4$ would predict 
exactly the opposite. 

A $Z^\prime$ with this characteristic
may in fact be a consequence of the $Z_2\times Z_2$
orbifold with standard embedding, which underlies the free
fermionic formulation for the following reason. Take, for
example, the case in which the anomalous $U(1)$ is
a combination which is embedded in $E_6$ and
is given by $U_A=U_1+U_2+U_3$ in the notation of Section 3.
The two anomaly free orthogonal combinations can be taken as
$U_1^\prime=U_1-U_2$ and $U_2^\prime=U_1+U_2-2U_3$.
The states of each generation from each sector $b_j$ have
charge $+1/2$ under $U(1)_j$ and are neutral with respect to
the other two. Consequently, $U_1^\prime$ produces 
charges which are equal in magnitude and opposite in sign
for two generation, whereas one generation is neutral
under it. This yields the same type of $Z^\prime$ that we examined
here and is strongly constrained by FCNC. On the
other hand $U_2^\prime$ is universal with respect
to two families and produces different charges
for the third family. This situation may, in fact,
be a unique consequence of the $Z_2\times Z_2$ orbifold
twisting, due to its cyclic permutation symmetry. From
Table 1 we see that, in fact, this type of charge
assignment is also frequently preserved in the three
generation models. What we argue is that if a $Z^\prime$ 
with universal couplings for the two light generations
and different couplings for the heavy generation
is observed in future experiments, it may be a key piece of
evidence for the $Z_2\times Z_2$ orbifold compactification.
In the case of a $Z^\prime$ with charges equal in
magnitude but opposite in sign for the first two
generations, we may expect it to be outside the reach
of forthcoming hadron colliders. However, if it is not too
far above their reach, we may expect novel FCNC
phenomena, and potentially new sources for CP violation.
We note that an additional $Z^\prime$ of the type that
we discussed here has also been advocated as playing a role
in suppressing proton decay in supersymmetric extensions
of the Standard Model \cite{pati}.
We also remark that very recently it has been suggested that
there exists evidence for a $Z^\prime$ with these 
characteristics in electroweak precision data \cite{erler}. 
All in all, nature may eventually prove to be kind for her
patient and obedient servants.

\section{Acknowledgments}
AF would like to heartily thank Misha Voloshin for
useful discussions on the matters of this paper and
physics in general.  
This work is supported in part
by DOE Grants No. DE--FG--0294ER40823 (AEF, TtV)
and DE--FG--0395ER40917 (GC,DVN).
\newpage
\appendix
\section{Quantum Number of FNY Massless Fields}

{\def\half{\frac{1}{2}}
\def\mhalf{-\frac{1}{2}}
\def\hfw{$\frac{1}{2}$}
\def\malf{-\frac{1}{2}}
\def\mfw{$-\frac{1}{2}$}
\def\sixth{\frac{1}{6}}
\def\third{\frac{1}{3}}
\def\mthird{-\frac{1}{3}}
\def\mbd{$\frac{2,-1}{3}$}
\def\mtd{$-\frac{1}{3}$}
\def\td{$\frac{1}{3}$}
\def\ttd{$\frac{2}{3}$}
\def\mttd{$-\frac{2}{3}$}
\def\mtwothird{\frac{2}{3}}
\def\mtwothird{-\frac{2}{3}}
\def\pmh{$\pm\half$}
\def\sutc{$SU(3)_C$}
\def\sutl{$SU(2)_L$}
\def\suth{$SU(3)_H$}
\def\sutl{$SU(2)_H$}
\def\sutn{$SU(2)^{'}_H$}
\def\UP#1{$U^{'}_{#1}$}
\def\U#1{$U_{#1}$}
\def\UC{$U_C$}
\def\UL{$U_L$}
\def\Ua{$U_A$}

\def\tb{$\bar{3}$}
\def\tbn{\bar{3}}

\def\T#1{$T_{#1}$}
\def\S#1{$S_{#1}$}
\def\H#1{$H_{#1}$}
\def\UR#1{$U_{#1}$}
\def\R#1{$R_{#1}$}
\def\b#1{$b_{#1}$}
\def\hv#1{$h_{#1}$}

\def\p#1{$\phi_{#1}$}
\def\pb#1{$\bar{\phi}_{#1}$}
\def\pp#1{$\phi^{'}_{#1}$}
\def\pbp#1{$\bar{\phi}^{'}_{#1}$}
\def\h#1{$h_{#1}$}
\def\hb#1{$\bar{h}_{#1}$}
\def\L#1{$L_{#1}$}
\def\ec#1{$e^c_{#1}$}
\def\Nc#1{$N^c_{#1}$}
\def\Q#1{$Q_{#1}$}
\def\dc#1{$d^c_{#1}$}
\def\uc#1{$u^c_{#1}$}
\def\Hs#1{$H^s_{#1}$}
\def\V#1{$V_{#1}$}
\def\Vs#1{$V^s_{#1}$}

\def\K#1{$K_{#1}$}

\begin{flushleft}
\begin{tabular}{|l||r|l|rrrrrrrrr|lccc|}
\hline 
\hline
state         &\U{E}&$(C,L)_Y$ &\Ua &\UC&\UL  &\UP{1}&\UP{2}&\UP{3}&\UP{4}&\UP{5}&\U{4}&$(3,2,2^{'})_H$&\U{7}&\U{H}&\U{9}\\
\hline
\hline
\Q{1}&  \mbd&  $(3,2)_{\sixth}$&       8&     2&     0&    -4&     2&     0&     0&   -24&     2&     (1,1,1)&   0&     0&     0\\ 
\Q{2}&  \mbd&  $(3,2)_{\sixth}$&      12&     2&     0&     2&   -10&     2&     2&    20&     0&     (1,1,1)&   0&     0&     0\\ 
\Q{3}&  \mbd&  $(3,2)_{\sixth}$&       8&     2&     0&     2&    14&    -2&     2&    32&     0&     (1,1,1)&   0&     0&     0\\ 
\hline
\dc{1}& \td&  $(\tbn,1)_{\third}$&    8&    -2&     4&    -4&     2&     0&     0&   -24&    -2&     (1,1,1)&   0&     0&     0\\ 
\dc{2}& \td&  $(\tbn,1)_{\third}$&    8&    -2&     4&     2&   -10&    -2&    -2&   -80&     0&     (1,1,1)&   0&     0&     0\\ 
\dc{3}& \td&  $(\tbn,1)_{\third}$&    4&    -2&     4&     2&    14&     2&    -2&   -68&     0&     (1,1,1)&   0&     0&     0\\

\hline
\uc{1}&\mttd&$(\tbn,1)_{\mtwothird}$&  8&    -2&    -4&    -4&     2&     0&     0&   -24&    -2&     (1,1,1)&   0&     0&     0\\ 
\uc{2}&\mttd&$(\tbn,1)_{\mtwothird}$& 12&    -2&    -4&     2&   -10&     2&     2&    20&     0&     (1,1,1)&   0&     0&     0\\ 
\uc{3}&\mttd&$(\tbn,1)_{\mtwothird}$&  8&    -2&    -4&     2&    14&    -2&     2&    32&     0&     (1,1,1)&   0&     0&     0\\ 
\hline
\ec{1}&   1& $(1,1)_{1}$&            8&     6&     4&    -4&     2&     0&     0&   -24&    -2&     (1,1,1)&   0&     0&     0\\ 
\ec{2}&   1& $(1,1)_{1}$&           12&     6&     4&     2&   -10&     2&     2&    20&     0&     (1,1,1)&   0&     0&     0\\ 
\ec{3}&   1& $(1,1)_{1}$&            8&     6&     4&     2&    14&    -2&     2&    32&     0&     (1,1,1)&   0&     0&     0\\ 
\hline
\Nc{1}&   0& $(1,1)_{0}$&            8&     6&    -4&    -4&     2&     0&     0&   -24&    -2&     (1,1,1)&   0&     0&     0\\ 
\Nc{2}&   0& $(1,1)_{0}$&            8&     6&    -4&     2&   -10&    -2&    -2&   -80&     0&     (1,1,1)&   0&     0&     0\\ 
\Nc{3}&   0& $(1,1)_{0}$&            4&     6&    -4&     2&    14&     2&    -2&   -68&     0&     (1,1,1)&   0&     0&     0\\ 
\hline
\L{1}& 0,-1& $(1,2)_{\mhalf}$&       8&    -6&     0&    -4&     2&     0&     0&   -24&     2&     (1,1,1)&   0&     0&     0\\
\L{2}& 0,-1& $(1,2)_{\mhalf}$&       8&    -6&     0&     2&   -10&    -2&    -2&   -80&     0&     (1,1,1)&   0&     0&     0\\ 
\L{3}& 0,-1& $(1,2)_{\mhalf}$&       4&    -6&     0&     2&    14&     2&    -2&   -68&     0&     (1,1,1)&   0&     0&     0\\ 
\hline
\h{1}& 0,-1& $(1,2)_{\mhalf}$&      16&     0&    -4&    -8&     4&     0&     0&   -48&     0&     (1,1,1)&   0&     0&     0\\ 
\h{2}& 0,-1& $(1,2)_{\mhalf}$&     -20&     0&    -4&    -4&    20&     0&     0&    60&     0&     (1,1,1)&   0&     0&     0\\ 
\h{3}& 0,-1& $(1,2)_{\mhalf}$&     -12&     0&    -4&    -4&   -28&     0&     0&    36&     0&     (1,1,1)&   0&     0&     0\\ 
\hb{1}&1, 0& $(1,2)_{ \half}$&     -16&     0&     4&     8&    -4&     0&     0&    48&     0&     (1,1,1)&   0&     0&     0\\
\hb{2}&1, 0& $(1,2)_{ \half}$&      20&     0&     4&     4&   -20&     0&     0&   -60&     0&     (1,1,1)&   0&     0&     0\\
\hb{3}&1, 0& $(1,2)_{ \half}$&      12&     0&     4&     4&    28&     0&     0&   -36&     0&     (1,1,1)&   0&     0&     0\\
\hline
\H{34}&1,0& $(1,2)_{ \half}$&       8&     3&     2&    -2&   -11&     2&    -4&    32&     0&     (1,1,1)&  -1&     3&     0\\ 
\H{41}&0,-1&$(1,2)_{\mhalf}$&       0&    -3&    -2&     2&   -13&    -2&    -4&    56&     0&     (1,1,1)&   1&    -3&     0\\ 
\hline
\hline
\end{tabular}
\end{flushleft}

\no Table 1: Gauge Charges of FNY three generation and Higgs sectors.
The names of the states appear in the first column, with the states' 
various charges appearing in the other columns.
The entries under $(C,L)_Y$ denote Standard Model charges, while
the entries under $(3,2,2')$ denote hidden sector $SU(3)_H\times SU(2)_H\times SU(2)^{'}_H$ charges. 
All $U(1)$ charges are multiplied by a factor of 4
relative to the definition in Eqs. (\ref{anomau1infny}) and (\ref{nonau1}).


               
\vfill\eject

\bigskip
\medskip

\def\bibiteml#1#2{ }
\bibliographystyle{unsrt}

\hfill\vfill\eject
\end{document}